\documentclass[sort&compress]{elsarticle}

\usepackage{amssymb}
\usepackage{graphicx}
\usepackage{bm}
\usepackage{xcolor}
\usepackage{amsmath}
\usepackage{indentfirst}
\usepackage[T1]{fontenc}
\usepackage[colorlinks,linkcolor=blue]{hyperref}
\usepackage{subcaption}

\journal{Physics of the Dark Universe}

\begin{document}
\begin{frontmatter}

\title{Induced Cosmological Anisotropies and CMB Anomalies by a non-Abelian Gauge-Gravity Interaction}

\author[label1,label2]{Bum-Hoon Lee}
\ead{bhl@sogang.ac.kr}
\author[label1,label2]{Hocheol Lee}
\ead{insaying@sogang.ac.kr}
\author[label1]{Wonwoo Lee}
\ead{warrior@sogang.ac.kr}
\author[label3,label4]{Nils A. Nilsson}
\ead{nilsson@ibs.re.kr}
\author[label5]{Somyadip Thakur}
\ead{somyadip@sogang.ac.kr}
\affiliation[label1]{organization={Center for Quantum Spacetime, Sogang University},
             city={Seoul},
             postcode={04107},
             country={Republic of Korea}}

\affiliation[label2]{organization={Department of Physics, Sogang University},
             city={Seoul},
             postcode={04107},
             country={Republic of Korea}}

\affiliation[label3]{organization={Cosmology, Gravity and Astroparticle Physics Group, Center for             Theoretical Physics of the Universe,
            Institute for Basic Science},
            city={Daejeon},
            postcode={34126},
            country={Republic of Korea}}

\affiliation[label4]{organization={SYRTE, Observatoire de Paris, Universit\'e PSL, CNRS, Sorbonne Universit\'e, LNE},
            addressline={61 avenue de l'Observatoire}, 
            city={Paris},
            postcode={75014},
            country={France}}

\affiliation[label5]{organization={Department of Physics, Hanyang University},
            city={Seoul},
            postcode={04763}, 
            country={Republic of Korea}}

\begin{abstract}
We present a non-abelian cousin of the model presented in \cite{Lee:2022rtz} which induces cosmological anisotropies on top of standard FLRW geometry. This is in some sense doing a cosmological mean field approximation, where the mean field cosmological model under consideration would be the standard FLRW, and the induced anisotropies are small perturbative corrections on top of it. Here we mostly focus on the non-abelian $SU(2)$ gauge fields coupled to the gravity to generate the anisotropies, which can be a viable model for the axion-like particle (ALP) dark sector. The induced anisotropies are consequences of the non-trivial back-reaction of the gauge fields on the gravity sector, and by a clever choice of the parametrization, one can generate the Bianchi model we have studied in this note. We also show that the anisotropies influence the Sachs-Wolfe effect and we discuss the implications.
\end{abstract}


\end{frontmatter}

\section{Introduction}
One the most successful and still a de-facto standard model of cosmology  has been the $\Lambda$ Cold Dark Matter (LCDM) model.
The basic building blocks of the model is the spatially flat, homogeneous and isotropic Friedmann-Lemaitre-Robertson-Walker (FLRW) geometry, which along with General Relativity and a positive cosmological constant is in excellent agreement with current cosmological observations. In this model, the only inhomogeneities are those of small perturbations in the early Universe, which together with the inflationary mechanism seeds formation of large-scale structure. Despite all this success, there are a number of issues currently being brought to light by new observations and the careful reanalysis of current data. The most glaring of these issues may be the {\it Hubble tension}, i.e. the $5\sigma$ discrepancy between the local (determined from the distance ladder) and cosmological (determined from the Cosmic Microwave Background (CMB)) value of the Hubble constant $H_0$ \cite{DiValentino:2021izs}. In light of this, a number of theoretical mechanisms have been explored in order to alleviate this tension, for example early dark energy \cite{Kamionkowski:2022pkx,Poulin:2023lkg}. The QCD axion is a compelling contender for beyond the Standard Model physics, since it is a natural candidate for Cold Dark Matter (CDM) and solves the strong CP problem \cite{Peccei:1977hh,Fukuda:2015ana}. Its pseudoscalar analogue in string theory is the axion-like particle (ALP), which is also interesting, as it introduces several important cosmological effects \cite{Choi:2022nlt}\footnote{See also our previous work on cosmology and ALP's \cite{Lee:2022rtz}.}; this is particularly relevant for cosmological tensions, as the axion field can introduce thermal friction in the early Universe. 

Recently, signals of {\it cosmic birefringence}, the parity-odd rotation of the polarization plane of E and B modes in the CMB, were reported, examined, and discussed in a series of papers \cite{Komatsu:2022nvu,Eskilt:2022wav, Eskilt:2022cff,Murai:2022zur,Minami:2019ruj,Nakatsuka:2022epj,Minami:2020odp,Diego-Palazuelos:2022cnh,Eskilt:2023nxm,Nilsson:2023sxz}, where the constraint on the polarization angle has now reached $\beta={0.342^\circ}^{+0.094^\circ}_{-0.091^\circ}$ ($1\sigma$) which is a non-zero signal at $3.6\sigma$. Such a signal can arise through several mechanisms, and it has been determined \cite{Eskilt:2022cff,Eskilt:2022wav} that the signal is consistent with being produced through an axion-photon coupling of the Chern-Simons $F_{\rm em}\widetilde{F}_{\rm em}$ type\footnote{Here, $F_{\rm em}$ is the electromagnetic field strength, with the subscript added to distinguish it from the $F$ used in Eq.~\ref{action1} and onwards.}, and such terms arise naturally in supergravity models (see for example \cite{freedman_van} for an introduction and review.). Apart from cosmic birefringence and the established Hubble parameter tension, the geometry of spacetime itself has recently come into question, with a number of observational probes reporting departures from the homogeneous and isotropic nature of FLRW. Hints of a quadrupole-octopole alignment in the CMB \cite{deOliveira-Costa:2003utu,Schwarz:2004gk}, apparent dipoles in the orientation of radio galaxy \cite{Schwarz:2004gk,Dolfi:2019wye}, as well as statistically significant signals of spatial variations of the fine-structure constant \cite{King:2012id} are challenging the cosmological Standard Model, which may need to be revised. There are also hints suggesting that certain combinations of datasets prefer a closed Universe \cite{DiValentino:2019qzk}.  

In a previous paper \cite{Lee:2022rtz} we investigated the scenario where the metric anisotropies in a Bianchi VII$_0$ model are induced by the non-trivial dynamics (back-reaction of the matter sector) of a $U(1)$ gauge field coupled to the metric, where we found that there exists an isotropic fixed point in the future, and that small anisotropies survive to the present time. Given these results, it is natural to generalise this to the case of a non-abelian gauge field which can be a viable candidate for the axion like dark sector particles. In this paper, we investigate the cosmological effects of an $SU(2)$ gauge field together with an ALP which can act as dark matter or dark energy, with particular focus on the geometry of the Universe.\footnote{We refer the readers for an exhaustive literature on related topics in \cite{Maleknejad_2011,maleknejad2013gaugeflation,Maleknejad_2012,Maleknejad_2018,Sheikh_Jabbari_2012,Maleknejad_2013} } We break the standard FLRW geometry by introducing a planar symmetry (or preferred symmetry axis) in the Universe in the form of a Bianchi VII$_0$ geometry, with a coupling between the SU(2) gauge field and the anisotropic metric functions. With the extra structure of the SU(2) gauge field, the $F\widetilde{F}$ term now affect the Einstein equations, which is not the case in the U(1) limit. Starting from a supergravity-inspired model with an SU(2) gauge field and an ALP, we introduce metric anisotropies and write down our equations of motion explicitly before solving our system using numerical integrators. We then derive the modified expression for the CMB temperature anisotropy in our geometry, and we compare it to current observations in order to put bounds on the anisotropic metric functions. We also compare our results with those we obtained in the U(1) sector in \cite{Lee:2022rtz}. In particle phenomenology, there are various kinds of non-abelian fields in addition to the simple $U(1)$, not only in the standard model but also in various models with hidden sectors. Physically, $U(1)$ and non-abelian fields have different physical behaviour and bearing on the phenomenology.

This paper is organized as follows: in Section~\ref{sec:GA} we introduce our model as well as all relevant anzätze for the gauge fields, and we write down the full set of equations of motion; in Section~\ref{sec:solmethods} we introduce the order-by-order solution scheme in preparation for the numerical solutions, and we outline our strategy for choosing initial conditions; in Section~\ref{sec:sols} we present and describe all solutions; in Section~\ref{sec:concl} we discuss our solutions and their implications in a broader context.

\section{Non-Abelian Gauge-Axion model} \label{sec:GA}

We begin this section by introducing the model. This is a non-abelian generalisation of the model that was earlier studied in \cite{Lee:2022rtz}. Phenomenological model building suggest that the axion-like particle, viz, the scalar particle is a viable candidate for the dark sector. In the abelian model, the scalar or the axion is minimally coupled to gravity and the coupling with the gauge sector drops off trivially. Generalising to a non-abelian model has certain advantages: for example, there is a non-minimal coupling of the gauge sector and the scalar sector, and the previously scalar field transforms as a pseudoscalar, providing us with a model which complements the phenomenological model for dark sector particles~\cite{Winch:2023qzl,OHare:2024nmr}. We will focus on the bosonic part of a supergravity-inspired model described by the action
\begin{equation}\label{action1}
	 S = \int d^4x \sqrt{-g}\Big[\frac{R-2\Lambda}{2\kappa}-\frac{1}{2}\nabla_\mu \phi \nabla^\mu \phi - V(\phi) - \frac{1}{4}F_{\mu\nu}^{~~a}F_a^{~\mu\nu}-\frac{\Theta \phi}{4}F_{\mu\nu}^{~~a}\widetilde{F}_a^{~\mu\nu}+ \mathcal{L}_{\rm PF}\Big] ,
\end{equation}
where $\kappa=8\pi G$ (which we set to unity from now on), $R$ is the  Ricci scalar, $\Lambda$ is the cosmological constant, $\phi$ is the pseudoscalar axion field, $\Theta$ is the axion decay constant, and  $\mathcal{L}_{\rm PF}$ is the canonical Lagrange density for a perfect fluid.  

The non-abelian field strength is given by $$F_{\mu\nu}^{~~a}=2 \partial_{[\mu}A_{\nu]}^a + g_A f^{a}_{~bc}  A_\mu^{~b}A_\nu^{~c},$$
where $a$ is the the colour index, $g_A$ is the $SU(2)$ coupling and $f^a_{~bc}$ are the structure constants. Also, $\widetilde{F}^a_{~\mu\nu}=\tfrac{1}{2}\epsilon^{\mu\nu\alpha\beta}F^a_{~\alpha\beta}$ is the dual field strength where $\epsilon^{\mu\nu\alpha\beta}$ is Levi-Civita tensor.\footnote{For detals we refer the readers to \ref{app:su2al}} The field $\phi$ can be thought of as a candidate for axionic dark matter and/or dark energy.
We note here that a stringent supergravity model would not allow us to have an explicit cosmological constant term in the action\footnote{There is a strong mismatch between the theoretically predicted and observed value of the cosmological constant \cite{Martin:2012bt}, and ideally, the cosmological constant should be small in a supergravity model due to supersymmetric constraints (cancellation between the fermionic and bosonic modes). However, the proposed supergravity models are far from such ideal cases. In order to provide a viable solution for the cosmological constant problem, one has to address lot of issues such as the SUSY breaking and vacuum energy problem, fine tuning issues, etc \cite{Weinberg:1988cp,Witten:2000zk}.} \cite{Freedman:2012zz}; however, for the present paper where we mostly study an effective cosmological model, such constraints coming from supergravity can be relaxed, and we present our action with an explicit cosmological constant term.

We consider the potential of the form\footnote{Note that here we have not considered temperature dependence in the potential, as presented in \cite{Choi:2022nlt}, the reason being the energy scale in which we are working in this paper, where the mass of the axion can be approximated with a constant value.}
\begin{equation}\label{pot:T}
V(\phi)=m_0^2 f^2\left(1-\cos{\frac{\phi}{f}}\right)
\end{equation}
\begin{figure}
\captionsetup[subfigure]{position=b}
     \centering
        \caption{The potential for the scalar field $\phi^{(0)}$.}
         \subcaptionbox{Potential for the scalar field $\phi^{(0)}$.\label{fig:phi0potfull}}{\includegraphics[width=0.48\linewidth]{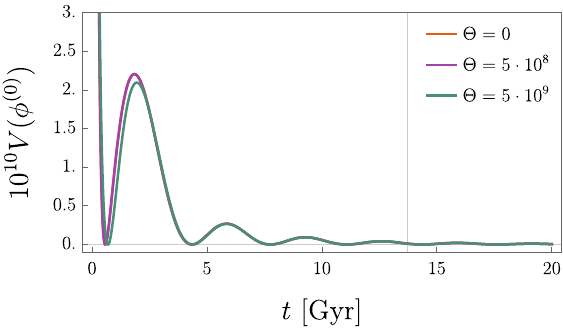}}
         \hfill
         \subcaptionbox{Potential for the scalar field $\phi^{(0)}$ at early times.\label{fig:phipotsmall}}{\includegraphics[width=0.48\linewidth]{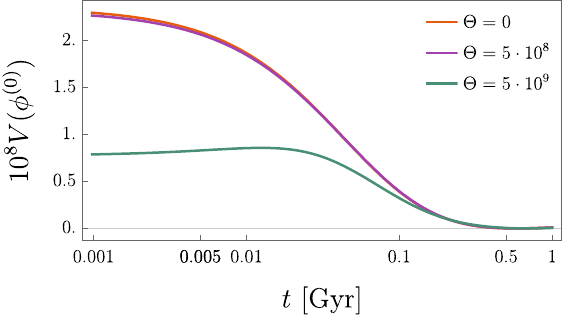}}
        \label{fig:phi0pot}
\end{figure}
Let us pause for a moment here to explain the nature of the potential as presented in Figure \ref{fig:phi0potfull} and \ref{fig:phipotsmall}. The potential is a function of the scalar field or the axion field $\phi$, which is determined as a solution of the coupled differential equation, i.e. the fully back-reacted solutions of the system. As the equation of motion governing $\phi$ goes from over-damped to critically damped behaviour as we increase the value of $\Theta$, the term which is dominant in the overdamped situation and becomes comparable to the contribution from the gauge fields in the critically damped situation, where the magnitude of the potential decreases, as seen in Figure \ref{fig:phipotsmall}.

We can write the equations of motion for \eqref{action1} as follows:\\
\underline{\it{The Einstein equations}}

\begin{equation}\label{eq:Eeqs}
    R_{\mu\nu}-\frac{1}{2}g_{\mu\nu}R +\Lambda g_{\mu\nu} =  \tilde{T}_{\mu\nu}^{\rm PF}+T_{\mu\nu}^{\rm AN}
\end{equation}
where $\tilde{T}_{\mu\nu}^{\rm PF}$ the stress-energy tensor for a perfect fluid.
We have written Eq.~\eqref{eq:Eeqs} in a familiar form by absorbing all scalar and gauge-field terms into the $T_{\mu\nu}^{\rm AN}$, which we call the anisotropic stress-energy tensor; it takes the form\footnote{Let us remind that the term $ \phi \sqrt{-g} \tilde{F}^{\mu\nu} F_{\mu \nu}=\frac{1}{2} \phi \epsilon^{\alpha \beta \mu \nu} F_{\alpha \beta} F_{\mu \nu}$ is independent of the metric and thus does not contribute the effective stress-energy tensor.}

\begin{equation}\label{anstr}
      T_{\mu\nu}^{\rm AN}=\nabla_\mu\phi\nabla_\nu\phi-\frac{1}{2}g_{\mu\nu}\nabla_\rho \phi \nabla^\rho \phi-g_{\mu\nu}V(\phi)-\frac{1}{4}g_{\mu\nu}F_{\rho\sigma}{}^{a} F_{a}{}^{\rho\sigma}+F_{\mu \rho}{}^{a}F_{a\nu}{}^{\rho}.
\end{equation}

\vspace{5mm}
\underline{\it{Equations of motion for $\phi$ and $A_\mu$}}
\begin{align}
    \label{eq:Phieq}0 \;&=\; \Box\phi - V^\prime(\phi)
    - \tfrac{1}{4}\Theta F_{\mu\nu}^{~~a}\widetilde{F}_a^{~\mu\nu} ,\\
    \label{eq:Aeq}0 \;&=\; \nabla^\nu\left(F_{\mu\nu}^{~~a}+\Theta\phi \widetilde{F}_{\mu\nu}^{~~a}\right) + g_A f^{a}{}_{bc} A^{\nu b}\left(F_{\mu\nu}^{~~c}+\Theta\phi \widetilde{F}_{\mu\nu}^{~~c}\right)
\end{align}

As in our previous work \cite{Lee:2022cyh}, we adopt the homogeneous and anisotropic Bianchi VII$_0$ metric, which can be parametrised as
\begin{equation}\label{eq:metric}
    ds^2 = -dt^2 + e^{2\alpha(t)}\left(e^{2\beta_1(t)}dx_1^2+e^{2\beta_2(t)}\left(dx_2^2+dx_3^2\right)\right),
\end{equation}
where $\alpha(t)$ and $\beta(t)$ are the isotropic and anisotropic scale factors, respectively. The factors of 2 have been chosen to coincide with the FLRW (isotropic, $\beta_i\rightarrow 0$) case, where $\dot{\alpha}(t)=\dot{a}(t)/a(t)$, $a(t)$ being the FLRW scale factor. 

\subsection{Gauge-field ansatz}
We would consider the expansion of the universe to be homogeneous but non-isotropic. We now choose to align the gauge field $A_\mu^a$ along the Killing vectors of the spacetime metric, and we can therefore parametrize the gauge field as
\begin{equation}
    A_i^a = \psi_i(t) e^a_i,
\end{equation}
where $e^a_i$ are the orthonormal frame fields related to \eqref{eq:metric}, which read
\begin{equation}
    e^a_i = e^{\alpha+\beta_i}\delta^a_i,
\end{equation}
leading to the gauge-field ansatz
\begin{equation}
    A_i^a = \left(e^{\alpha+\beta_1}\psi_1\delta^a_1, \; e^{\alpha+\beta_2}\psi_2\delta^a_2, \; e^{\alpha+\beta_3}\psi_3\delta^a_3\right),
\end{equation}

throughout the rest part of this paper we will work in the temporal gauge.
With these expressions in mind, we write the components of the SU(2) field strength as
\begin{equation}
    \begin{aligned}
    F_{0i}^a \; &= \; \partial_0A^a_i = \partial_0\left(e^{\alpha+\beta_i}\psi_i\right)\delta_i^a \\
    F_{ij}^a \; &= \; -g f^{a}_{bc}A_i^bA_j^c = -gf^{a}_{ij}\left(e^{2\alpha+\beta_i+\beta_j}\right)\psi_i\psi_j,
  \end{aligned}
\end{equation}
which we now use to rewrite the equations of motion for $\phi$, $A_\mu^a$, and $g^{\mu\nu}$ where $\psi_4 \equiv \psi_1$, $\psi_5 \equiv \psi_2$, $\beta_4 \equiv \beta_1$ and $\beta_5 \equiv \beta_2$:
\begin{equation}\label{eq:eomsphiA}
\begin{aligned}
    \phi: \; 0 \; &= \; \ddot{\phi} + \left( 3 \dot{\alpha} + \sum_{i=1}^3 \dot{\beta}_i \right) \dot{\phi} + V'(\phi) \\&+ g_A \Theta \Big[ \big( 3 \dot{\alpha} + \sum_{i=1}^3 \dot{\beta}_i \big) \psi_1 \psi_2 \psi_3 + \sum_{i=1}^3 \dot{\psi}_i \psi_{i+1} \psi_{i+2} \Big],
    \\
    A_{i}^a: \; 0 \; &= \; \ddot{\psi}_i + \left( 3 \dot{\alpha} + \sum_{i=1}^3 \dot{\beta}_i \right) \dot{\psi}_i - g_A \Theta \psi_{i+1} \psi_{i+2} \dot{\phi} + \Big[ \ddot{\alpha} + \ddot{\beta}_i \\& + \left( \dot{\alpha} + \dot{\beta}_i \right) \left( 2 \dot{\alpha} + \dot{\beta}_{i+1} + \dot{\beta}_{i+2} \right)  + g_A^2 \left( \psi_{i+1}^2 + \psi_{i+2}^2 \right) \Big] \psi_i,
    \\
    g^{tt}: \; 0 \; &= \; 3 \dot{\alpha}^2 + 2 \dot{\alpha} \sum_{i=1}^3 \dot{\beta}_i + \sum_{i=1}^3 \dot{\beta}_i \dot{\beta}_{i+1} - \frac{\kappa}{2} \Bigg\{ \sum_{i=1}^3 \left[ \dot{\psi}_i + \left( \dot{\alpha} + \dot{\beta}_i \right) \psi_i \right]^2  \\&  + g_A^2 \sum_{i=1}^3 \left( \psi_i^2 \psi_{i+1}^2 \right) + \dot{\phi}^2 + 2 V(\phi) + 2 \rho_{\rm PF} \Bigg\},
    \\
    g^{ii}: \; 0 \; &= \; 2 \ddot{\alpha} + \ddot{\beta}_{i+1} + \ddot{\beta}_{i+2} + 3 \dot{\alpha}^2 + 3 \dot{\alpha} \left( \dot{\beta}_{i+1} + \dot{\beta}_{i+2} \right) + \left( \dot{\beta}_{i+1} + \dot{\beta}_{i+2} \right)^2 \\& - \dot{\beta}_{i+1} \dot{\beta}_{i+2} + \frac{\kappa}{2} \Big\{ \dot{\phi}^2 - 2 V(\phi) - \left[ \dot{\psi}_i + \left( \dot{\alpha} + \dot{\beta}_i \right) \psi_i \right]^2 + \big[ \dot{\psi}_{i+1} \\& + \left( \dot{\alpha} + \dot{\beta}_{i+1} \right) \psi_{i+1} \big]^2 + \left[ \dot{\psi}_{i+2} + \left( \dot{\alpha} + \dot{\beta}_{i+2} \right) \psi_{i+2} \right]^2 \\& + g_A^2 \left( \psi_i^2 \psi_{i+1}^2 + \psi_i^2 \psi_{i+2}^2 - \psi_{i+1}^2 \psi_{i+2}^2 \right) + 2 p_{\rm PF} \Big\}.
\end{aligned}
\end{equation}

\section{Solving the equations of motion}\label{sec:solmethods}
In order to make the equations of motion more tractable, we reparametrize the gauge field $A_\mu^1$\footnote{We are working in the diagonal basis of the $SU(2)$, where the equations of motion govering the $A^a_\mu$ are the same for $a=1,2,3$ (the three color factors). We just present here the parametrization of one of the gauge field and the rest of the parametrizations are same. In the final results we have taken into account for all the gauge fields. For more details we refer the readers to   \ref{app:su2al}} by introducing two fiducial scalar fields which we call $\sigma(t)$ and $\psi(t)$; we then write $\psi_i(t)$ as
\begin{equation}
    \begin{aligned}
        \psi_1(t) = \frac{\psi(t)}{\sigma(t)^2}, \quad \psi_2(t) = \psi_3(t) = \sigma(t)\psi(t),
    \end{aligned}
\end{equation}
which preserves the number of degrees of freedom in the system and will be useful when isolating the FLRW limit of the equations of motion. Given the above redefinition, we observe that the isotropic limit is given by
\begin{equation}
    \dot{\beta}_i(t) = 0, \quad \psi(t) = 0, \quad \sigma(t) = \pm 1.
\end{equation}

In order to solve the equations of motion, we follow the approach in our previous work \cite{Lee:2022rtz} and introduce the following perturbative scheme
\begin{enumerate}
    \item Expand all scalar fields around their isotropic fixed points and retain only terms linear in the perturbation parameter $\epsilon$ as $X(t) = X^{(0)}(t) + \epsilon X^{(1)}(t)$, where $\epsilon$-order terms represent the anisotropic contribution.
    \item Solve the system at zeroth order ($\epsilon \to 0$), which corresponds to the homogeneous and isotropic solution.
    \item Use the zeroth-order solutions as seeds in the first-order equations to find the full solutions.
\end{enumerate}
In doing this, we make the implicit assumption that anisotropies are small (backed by observation), and we can treat the full equations as small anisotropies on top of an FLRW background; as such, we use an FLRW seed solution in the order-by-order scheme which simplifies the coupled system of equations. With the order-by-order scheme above, we define the expansions around the homogeneous and isotropic limit as
\begin{equation}
\begin{aligned}
    & \alpha(t) = \alpha^{(0)}(t), && \beta_2(t) = \epsilon\beta_2^{(1)}(t), &&& \beta_3(t) = \epsilon\beta_3^{(1)}(t) \\ & \psi(t) = \psi^{(0)}(t) + \epsilon\psi^{(1)}(t), && \phi(t) = \phi^{(0)}(t)+\epsilon\phi^{(1)}(t), &&& \sigma(t) = \pm 1 + \epsilon\sigma^{(1)}(t),
\end{aligned}
\end{equation}
where we have incorporated the homogeneous and isotropic limit by setting $\sigma^{(0)}(t) = 0$, $\beta_i^{(0)}=0$, and fixed the gauge by setting $\alpha^{(1)}(t)=0$\footnote{See Appendix A in \cite{Lee:2022rtz}} . 

In Eq.~(\ref{eq:Eeqs}) we introduced the perfect-fluid stress-energy tensor $\tilde{T}_{\mu\nu}^{\rm PF}$, where the tilde denotes the absorption of the cosmological constant, i.e. $T_{\mu\nu}^{\rm PF}=\tilde{T}_{\mu\nu}^{\rm PF}-\Lambda g_{\mu\nu}$ which now reads
\begin{equation}
    T_{\mu\nu}^{\rm PF} =
    \begin{pmatrix}
        \rho & 0 & 0 & 0 \\
        0 &&& \\
        0 & & g_{ij}p & \\
        0 &&&
    \end{pmatrix}
\end{equation}
where $\rho$ is the energy density and $p$ is the pressure of the cosmic fluid. In the spatially flat limit, the zeroth-order stress-energy tensor reads
\begin{equation}
    T_{\mu\nu}^{\rm{PF}, (0)}=
    \begin{cases}
        3H_0^2(\Omega_r^0e^{-4\alpha}+\Omega_m^0e^{-3\alpha}+\Omega_\Lambda^0), &(\mu=\nu=0) \\
        3H_0^2(\tfrac{1}{3}\Omega_r^0e^{-4\alpha}-\Omega_\Lambda^0)e^{2\alpha}, &(\mu=\nu=i),
    \end{cases}
\end{equation}
where $\Omega_X^0$ are the standard fractional energy densities for matter ($m$), radiation ($r$), and cosmological constant ($\Lambda$) as measured today. The first-order expression (as well as the full equations of motion) is presented in \ref{app:pert}.

\subsection{Zeroth-order equations}\label{sec:zeroth}
We begin by presenting the zeroth-order equations and discuss their properties. Starting with the scalar field $\phi(t)$, we expand around the isotropic fixed point and take the limit $\epsilon \to 0$, after which the equation of motion (\ref{eq:eomsphiA}) reads
\begin{equation}\label{eqn:zeroth}
    0 = \ddot{\phi}^{(0)} + 3\dot{\alpha}^{(0)}\left(g_A\Theta(\psi^{(0)})^3+\dot{\phi}^{(0)}\right) + 3g_A\Theta (\psi^{(0)})^2\dot{\psi}^{(0)}+V'(\phi^{(0)}),
\end{equation}
where we see that the SU(2)-induced coupling contributes also at zeroth order. At zeroth order, the components of the gauge field potential $A_\mu^a$ are identical and read
\begin{equation}\label{eqn:1storder}
    0 = \ddot{\psi}^{(0)}+3\dot{\alpha}\dot{\psi}^{(0)}+\psi^{(0)}\left(\ddot{\alpha}+2\dot{\alpha}^2\right)+2g_A^2(\psi^{(0)})^3-g_A\Theta(\psi^{(0)})^2\dot{\phi}^{(0)}.
\end{equation}
The zeroth-order Einstein equations read
\begin{equation}
    \begin{aligned}
        0 \; &= \; 3\dot{\alpha}^2 - \kappa V(\phi^{(0)})-\kappa \rho -\frac{\kappa}{2}(\dot{\phi}^{(0)})^2 -\frac{3\kappa}{2}\Big(g_A^2(\psi^{(0)})^4 +(\psi^{(0)})^2\dot{\alpha}^2 \\& +2\psi^{(0)}\dot{\psi}^{(0)}\dot{\alpha} +(\dot{\psi}^{(0)})^2\Big), \quad (\mu=\nu=0), \\
        0 \; &= \; 3\dot{\alpha}^2+2\ddot{\alpha} + \frac{\kappa}{2}\Big(2p-2V(\phi^{(0)})+g_A^2(\psi^{(0)})^4+(\psi^{(0)})^2\dot{\alpha}^2+(\dot{\phi}^{(0)})^2\\&  +2\psi^{(0)}\dot{\psi}^{(0)}\dot{\alpha}+(\dot{\psi}^{(0)})^2\Big), \quad (\mu=\nu=i).
    \end{aligned}
\end{equation}

\subsection{First order}
At first order, the equation for the scalar field $\phi^{(1)}$ reads\footnote{From now on, we will enclose $\epsilon$-order quantities in square brackets.}
\begin{equation}
\begin{aligned}
    0&=\ddot{\phi}^{(0)}+3 g_A  \Theta  (\psi^{(0)})^3 \dot{\alpha}^{(0)}+3 \dot{\alpha}^{(0)} \dot{\phi}^{(0)}+3 g_A  \Theta  (\psi^{(0)})^2 \dot{\psi}^{(0)}+V^\prime(\phi^{(0)})
    \\&
     +\epsilon  \Big[\ddot{\phi}^{(1)}+3 \dot{\alpha}^{(0)} \dot{\phi}^{(1)}+\dot{\phi}^{(0)} \left(\dot{\beta}^{(1)}+2 \dot{\beta}_2^{(1)}\right)+3 g_A  \Theta  (\psi^{(0)})^2 \dot{\psi}^{(1)} \\& +6 g_A  \Theta  \psi^{(0)} \psi^{(1)} \dot{\psi}^{(0)}
     +9 g_A  \Theta  (\psi^{(0)})^2 \psi^{(1)} \dot{\alpha}^{(0)}  +\phi^{(1)} V^{\prime\prime}(\phi^{(0)})\\& +g_A  \Theta  (\psi^{(0)})^3 \left(\dot{\beta}^{(1)}+2 \dot{\beta}_2^{(1)}\right)\Big]
\end{aligned}
\end{equation}
From the first-order equations of motion for the gauge field $A_\mu^a$ we have that only the ``diagonal'' components are non-zero, i.e. $A_i^i$ (no sum), and that $A_2^2=A_3^3$; they are lengthy and we display them in \ref{app:pert}.

The first Friedmann equation ($\mu=\nu=0$ component of the Einstein equations) reads
\begin{equation}
\begin{aligned}
    0&=3 (\dot{\alpha}^{(0)})^2-\frac{\kappa}{2} \Big( (\dot{\phi}^{(0)})^2 + 3 (\psi^{(0)})^2 (\dot{\alpha}^{(0)})^2 + 3 (\dot{\psi}^{(0)})^2 + 6 \psi^{(0)} \dot{\alpha}^{(0)} \dot{\psi}^{(0)} \\& + 3 g_A^2 (\psi^{(0)})^4 + 2 \rho +2 V(\phi^{(0)})\Big) +\epsilon  \Big\{2 \dot{\alpha}^{(0)} \left(\dot{\beta}_1^{(1)}+2 \dot{\beta}_2^{(1)}\right)  \\& - \kappa \Big[ \dot{\phi}^{(0)} \dot{\phi}^{(1)} + 6 g_A^2 (\psi^{(0)})^3 \psi^{(1)} + \phi^{(1)} V^\prime(\phi^{(0)})
    \\& + 3 \left( \dot{\alpha}^{(0)} \psi^{(0)} + \dot{\psi}^{(0)} \right) \left(\psi^{(1)} \dot{\alpha}^{(0)}+\dot{\psi}^{(1)}\right) \\& + \Big( (\psi^{(0)})^2 \dot{\alpha}^{(0)} - \psi^{(0)} \dot{\psi}^{(0)} \Big) \left(\dot{\beta}_1^{(1)}+2 \dot{\beta}_2^{(1)}\right) \Big]\Big\}
\end{aligned}
\end{equation}

The rest of the decomposed Einstein equations are quite lengthy at first order, and we display them in \ref{app:pert}

\subsection{Numerical setup and boundary conditions  }\label{sec:numsol}
We solve the full system of equations for the Einstein, gauge field, and scalar field parts order-by-order according to the prescription in Section \ref{sec:solmethods}, where we choose initial conditions in a consistent way through the relevant equations of motion, since all the variables are coupled. We call the ones we are free to choose ``primary'' initial conditions, which we show in Table \ref{tab:initconds}; we follow the same procedure as in Appendix D of our paper \cite{Lee:2022rtz}. 
\begin{table}[h]
\begin{center}
\begin{tabular}{c c c c}
Zeroth order & & & \\
\hline
\hline \\
     $a(t_f)=1.6$ & $\phi^{(0)}(t_f)=10^{-6}$ & $\dot{\phi}^{(0)}(t_f)=-10^{-6}$ & $\psi^{(0)}(t_f)=10^{-6}$ \\
     & & & \\
First order & & &\\
\hline
\hline \\
    $\phi^{(1)}(t_f)=10^{-6}$ & $\dot{\phi}^{(1)}(t_f)=-10^{-6}$ & $\psi^{(1)}(t_f)=10^{-6}$ & $\dot{\psi}^{(1)}(t_f)=-10^{-6}$ \\
    $\sigma^{(1)}(t_f)=10^{-3}$ & $\dot{\sigma}^{(1)}(t_f)=-10^{-3}$ &
    $\beta_1^{(1)}(t_f)=10^{-6}$ & $\dot{\beta}_1^{(1)}(t_f)=-10^{-6}$ \\
    $\beta_2^{(1)}(t_f)=-10^{-6}$ & & & \\
\end{tabular}
\caption{Boundary conditions set at $t_f=20$ Gyr, the FLRW fixed point.}
\label{tab:initconds}
\end{center}
\end{table}
We also choose the model parameters as follows:
\begin{equation}
m_0 = 1, \quad f = 1, \quad g_A = 1, \quad \Theta \sim 10^9,
\end{equation}
as well as the best-fit cosmological parameters from the {\it Planck} 2018 data release \\(TT,TE,EE+lowE+lensing+BAO): $\Omega_m^0 = 0.3111, \Omega_\Lambda^0=0.6889, \Omega_r^0=9.18\cdot 10^{-5}$ \cite{Planck:2018vyg}.
The initial values presented in Table\ref{tab:initconds} can be justified as follows: 
at early times, the anisotropies cannot be too large, as that would (for example) cause the CMB temperature quadrupole should to deviate too far from the measured value (see also Section \ref{sec:CMB}); at late times the solution should respect the cosmic no-hair theorem but still allow for a small amount of anisotropy to survive at the present time. With this in mind, we set the initial conditions to those in Table \ref{tab:initconds}.

It is worthwhile to note at this point that the axion decay constant or the axion gauge field coupling constant $\Theta$ have a magnitude which is of the order of $\sim 10^9$. This seems like a very large value, but we can intuitively give a rough order-of-magnitude explanation for this: Eq.~\eqref{eqn:zeroth} is roughly the equation for a damped harmonic oscillator; there are two competing terms in the equation, one coming from the potential and the other (the damping term) coming from the non-abelian nature of the the axion-gauge field coupling. Depending on the magnitude of $\Theta$ and the potential $V(\phi)$, we can have the following scenarios:
\begin{itemize}
    \item Overdamped expansion,
    \item Critically damped expansion,
    \item Under-damped expansion.
\end{itemize}
When $\Theta=0$, the system is overdamped, which can be seen from the orange curve in Figure~\ref{fig:phi1}. As we increase the magnitude of $\Theta$ such that we cross the region from overdamped $\to$ criticlly damped $\to$ slightly underdamped, the non-abelian terms and the contribution from the potential becomes comparable, and the effect of the non-abelian contribution is evident. Therefore, we choose values of $\Theta$ to capture the behaviour of these three regions of the solution space. 

From the zeroth-order Einstein equations, we solve for the isotropic scale factor $\alpha^{(0)}$, where we impose boundary conditions at the isotropic fixed point and solve for the evolution. From the zeroth-order scalar and gauge-field equations, we can find $\phi^{(0)}$ and $\psi^{(0)}$, respectively. Since our equations of motion contain both growing and decaying modes, we choose boundary conditions such that there exists a homogeneous and isotropic fixed point in the future, in order to satisfy current observations as well as the cosmic no-hair theorem \cite{Wald:1983ky}; in our solutions everything settles down to the FLRW Universe. 

\section{Solutions and Applications}\label{sec:sols}
We solve the full system of order-by-order equations for scalar, vector, and tensor contributions numerically and present the relevant solutions here. Qualitatively, the solutions indicate that the contributions from the anisotropies and the axionic potential $V(\phi)$ were large in the early Universe before decaying exponentially and flowing to the homogeneous and isotropic fixed point corresponding to pure FLRW. 

\begin{figure}
\captionsetup[subfigure]{position=b}
     \centering
        \caption{The isotropic scale factor $\alpha^{(0)}$.}
         \subcaptionbox{The isotropic scale factor $\alpha^{(0)}$ for three different values of the SU(2) coupling constant $\Theta$\label{fig:alpha0}}{\includegraphics[width=0.48\linewidth]{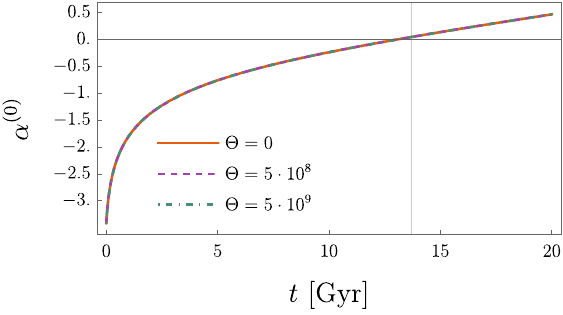}}
         \hfill
         \subcaptionbox{Ratio of the isotropic scale factor $\alpha^{(0)}$ for three different values of the SU(2) coupling constant $\Theta$\label{fig:alpha0ratio}}{\includegraphics[width=0.48\linewidth]{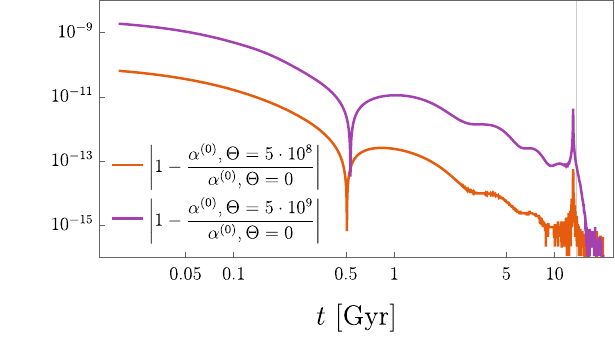}}
        \label{fig:alpha0both}
\end{figure}

\begin{figure}
\captionsetup[subfigure]{position=b}
     \centering
        \caption{Solutions for for the scalar field $\phi$.}
         \subcaptionbox{The scalar field $\phi^{(0)}$ for three different values of the SU(2) coupling constant $\Theta$\label{fig:phi0}}{\includegraphics[width=0.48\linewidth]{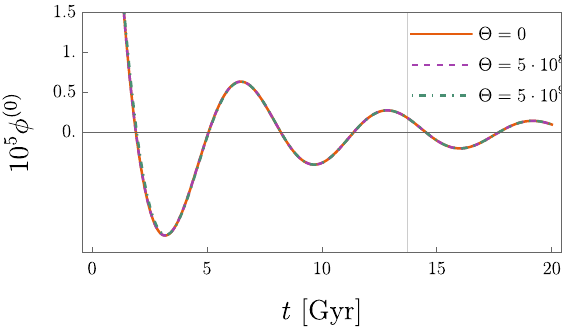}}
         \hfill
         \subcaptionbox{The scalar field $\phi^{(1)}$ for three different values of the SU(2) coupling constant $\Theta$\label{fig:phi1}}{\includegraphics[width=0.48\linewidth]{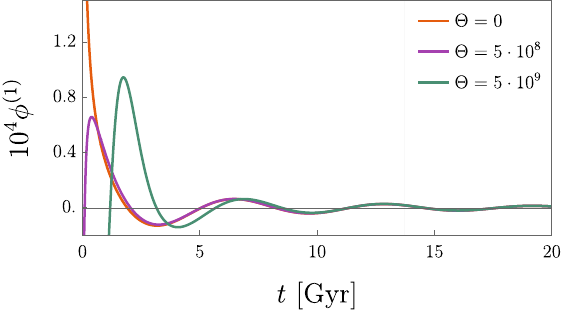}}
        \label{fig:phi}
\end{figure}

In Figure \ref{fig:alpha0} we see that the neither the anisotropy nor the non-abelian nature of the universe has any effect on the isotropic scale factor, which is in line with our expectations. As can be seen in Figure \ref{fig:alpha0}, the value at the present time $H_0^{-1}=13.787$ Gyr is slightly different than the standard choice in $\Lambda$CDM; this is an artefact of imposing the initial conditions at $t=20$ Gyr. Next we study the scalar field $\phi^0$, which can be seen in Figure \ref{fig:phi0}; here, we can see the effect of both the non-abelian contributions and the axionic potential chosen, and we see that from $t=0$ to $t\sim 2$ Gyr, increasing the coupling constant $\Theta$ increases the value of $\phi^{(0)}$. We also show the ratio between the scale factors for different values of $\Theta$ in Figure~\ref{fig:alpha0ratio}, where we see that the difference is always smaller than $10^{-9}$\footnote{Since the difference from the $\Lambda$CDM in $\bar{H}$ is so small, our model generally lies within the error bars of local distance measurements, which report errors on the order of $10^{-2}$ (see for example \cite{Riess:2021jrx}).} and that numerical noise dominates after $5$ Gyr. 

\begin{figure}[h]
    \centering
     \caption{The equation of state parameter for $\phi^{(0)}$ for three different values of the SU(2) coupling constant $\Theta$}
    \includegraphics[scale=0.7]{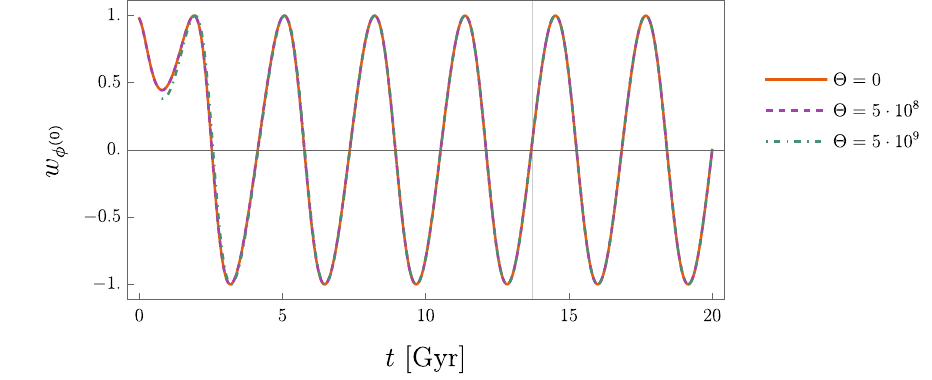}
    \label{fig:wphi0}
\end{figure}

Overall, the solution behaves as a damped harmonic oscillator, and the changes in amplitudes survives well past the present day. As this scalar may act as dark matter or dark energy, we also study its equation of state
\begin{equation}
    w_{\phi^{(0)}} = \frac{p_{\phi^{(0)}}}{\rho_{\phi^{(0)}}}, 
\end{equation}
which we plot in Figure \ref{fig:wphi0}. Here, we clearly see a smooth, non-damped oscillation between $+1$ and $-1$, i.e. between {\it stiff matter} and a {\it cosmological constant} with a period on the order of a few Gyr. As in the case of $\phi^{(0)}$, the non-abelian nature can only be seen at very early times. An interesting feature in the equation of state is the existence of a kink starting at $t\sim0.5$ Gyr.

\begin{figure}
\captionsetup[subfigure]{position=b}
     \centering
        \caption{Solutions for for the scalar field $\psi$.}
         \subcaptionbox{The scalar field $\psi^{(0)}$ for three different values of the SU(2) coupling constant $\Theta$\label{fig:psi0}}{\includegraphics[width=0.48\linewidth]{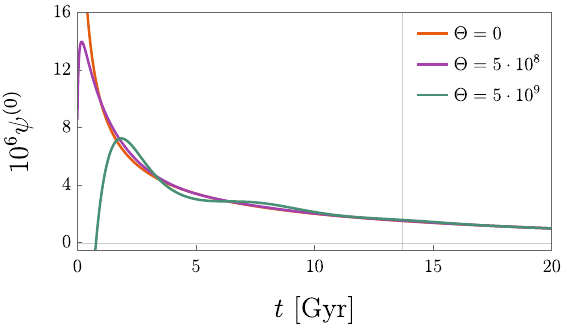}}
         \hfill
         \subcaptionbox{The scalar field $\psi^{(1)}$ for three different values of the SU(2) coupling constant $\Theta$\label{fig:psi1}}{\includegraphics[width=0.48\linewidth]{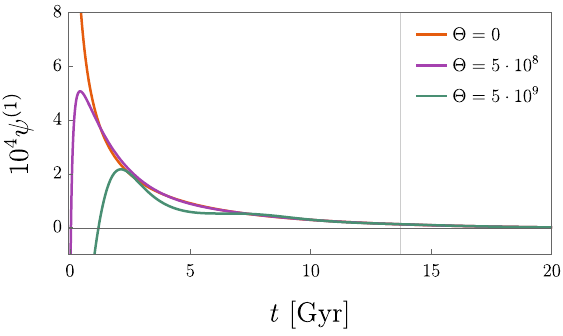}}
        \label{fig:psi}
\end{figure}

At first order in the anisotropies, we find the solutions to the scalar field $\phi^{(1)}$ to be that of a damped oscillator at late times (much like the solution for $\phi^{(0)}$) but with a significantly smaller amplitude. At early times, these oscillations becomes sizeable, with a peak which depends on the value of $\Theta$, before taking on negative values as $t \to 0$. As we increase the value of $\Theta$, the maximum moves to larger values of $t$, as can be seen in Figure~\ref{fig:phi1}\footnote{Compared to the magnitude of $\phi^{(0)}$, the first-order solution is very small and it should be multiplied further with the expansion parameter $\epsilon$ when constructing the full solution $\phi = \phi^{(0)}+\epsilon \phi^{(1)}$. For this reason, we do not include plots of the full solution, as it would be difficult to discern the difference.}.
The gauge-field component $\psi$ shows similar behaviour at both zeroth and first order, with the solutions for smaller $\Theta$ diverging at early times, with a maximum appearing as $\Theta$ is increased along with the oscillatory behaviour arising due to the axionic potential. At late times, the oscillations are significantly damped, leading to an exponentially decaying solutions for all values of $\Theta$. At zeroth order, a peak appears at lower values of $\Theta$ compared to first order, which can be seen in Figures~\ref{fig:psi0} and \ref{fig:psi1}.

\begin{figure}[h]
    \centering
    \caption{The scalar field $\sigma^{(1)}$ for three different values of the SU(2) coupling constant $\Theta$}
    \includegraphics[scale=0.8]{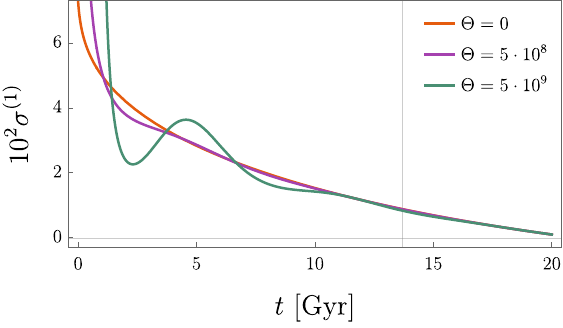}
    \label{fig:sigma1}
\end{figure}

The second scalar component of the gauge field is $\sigma^{(1)}$, which assumes a similar profile, albeit without turning points towards negative values. Instead, the oscillatory nature persists even at lower values of $\Theta$, and dominates the solution as $\Theta$ increases, with small oscillations still visible at the present time, as can be seen in Figure~\ref{fig:sigma1}.

\begin{figure}[h]
    \caption{The anisotropic scale factors $\beta_{1,2}^{(1)}$ for three different values of the SU(2) coupling constant $\Theta$}
    \hspace{-10mm}\includegraphics[scale=0.8]{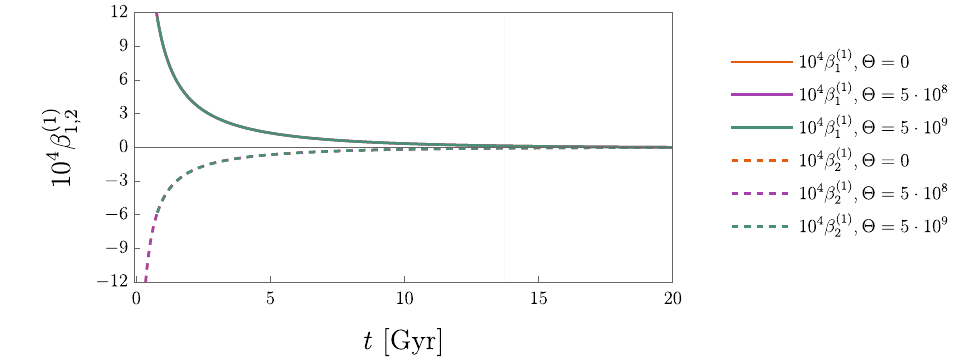}
    \label{fig:beta12}
\end{figure}

\begin{figure}
\captionsetup[subfigure]{position=b}
     \centering
        \caption{Ratios of the anisotropic scale factor $\beta_i^{(1)}$.}
         \subcaptionbox{Ratios of the anisotropic scale factor $\beta_1^{(1)}$ for three different values of the SU(2) coupling constant $\Theta$\label{fig:beta1ratio}}{\includegraphics[width=0.48\linewidth]{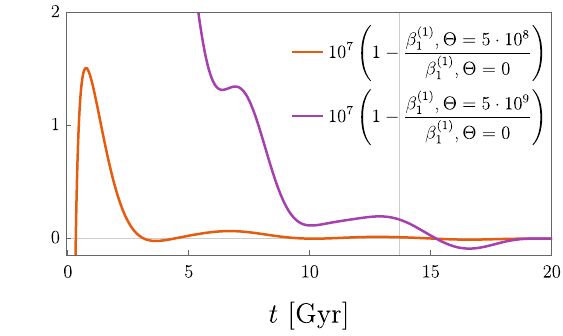}}
         \hfill
         \subcaptionbox{Ratios of the anisotropic scale factor $\beta_2^{(1)}$ for three different values of the SU(2) coupling constant $\Theta$\label{fig:beta2ratio}}{\includegraphics[width=0.48\linewidth]{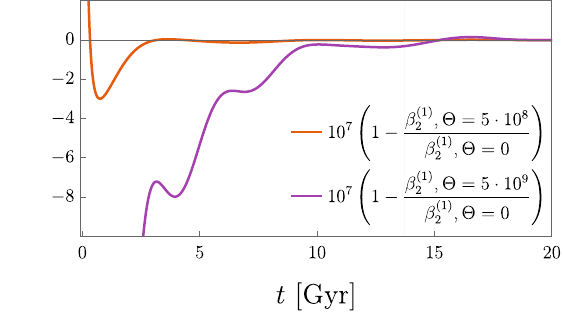}}
        \label{fig:betaratio}
\end{figure}

Next, we turn to the anisotropic scale factors $\beta_1^{(1)}$ and $\beta_2^{(1)}$, which are displayed in Figure~\ref{fig:beta12} for different values of $\Theta$. Both $\beta$'s appear to exhibit a smooth exponential falloff and have an approximate mirror symmetry $\beta_1^{(1)}+\beta_2^{(1)}\approx 0$, which does not seem to depend on the value of $\Theta$.
We notice, however, that when plotting the ratio between the same anisotropic scale factor for different values of $\Theta$, a clear oscillatory behaviour appears, as can be seen in Figure~\ref{fig:beta1ratio} (for $\beta_1^{(1)}$) and \ref{fig:beta2ratio} (for $\beta_2^{(1)}$), which reveal subleading oscillations on the order of $10^{-7}$. We also observe that the decay of the SU(2) features occurs at later time for larger values of $\Theta$. 

\begin{figure}
\captionsetup[subfigure]{position=b}
     \centering
        \caption{The average Hubble parameter $\bar{H}$.}
         \subcaptionbox{The average Hubble parameter $\bar{H}$ for three different values of the SU(2) coupling constant $\Theta$\label{fig:hbar}}{\includegraphics[width=0.48\linewidth]{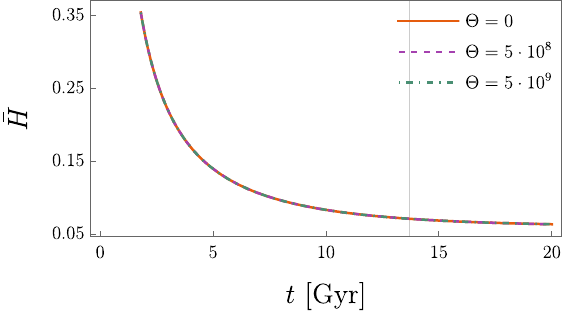}}
         \hfill
         \subcaptionbox{Ratios of the Hubble parameter $\bar{h}$ for three different values of the SU(2) coupling constant $\Theta$\label{fig:hbarratio}}{\includegraphics[width=0.48\linewidth]{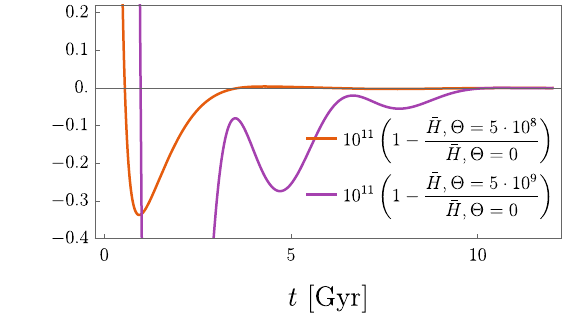}}
        \label{fig:Haverage}
\end{figure}

A common discriminator when working with anisotropic cosmology is the average Hubble parameter, which we denote with an overbar on $H$, and which in our case can be written as 
\begin{equation}\label{eq:Hbar}
    \bar{H}(t) \equiv \frac{1}{3}\left(3\dot{\alpha}^{(0)}+\epsilon \beta_1^{(1)}+2\epsilon\beta_2^{(1)}\right).
\end{equation}
We plot the behaviour of $\bar{H}$ in Figure~\ref{fig:hbar} where we see that the difference between different values of $\Theta$ cannot be observed, although there are hints of oscillations appearing at late times. Instead, we normalise $\bar{H}$ with its corresponding behaviour when $\Theta=0$ (as in Figures~\ref{fig:beta1ratio},\ref{fig:beta2ratio} for $\beta_i^{(1)}$), where we observe that a deviation from the $\Theta=0$ case appears on the order of $10^{-11}$, with oscillations showing for $\Theta=5\cdot10^9$; this can be seen in Figure~\ref{fig:hbarratio}.

\subsection{Cosmic microwave background temperature anisotropies}\label{sec:CMB}
In modern cosmology the Cosmic Microwave Background (CMB) remains one of the most powerful and important  tool to study early-Universe cosmology; the $\Lambda$CDM model was confirmed at a high level of accuracy in the final data release by the {\it Planck} collaboration (see for example \cite{Planck:2018lkk,Planck:2018vyg, Planck:2018lbu}). Nevertheless, several anomalies persist in the {\it Planck} data, particularly at large scales \cite{Schwarz:2015cma, Cea:2019gnu}, the most prominent of which is the quadrupole temperature correlation which remains heavily suppressed as compared to the best fit $\Lambda$CDM model. One possible solution to this problem is the introduction of a metric with a $\mathbb{R}^2\times\mathbb{R}$ spatial symmetry (planar symmetry) rather than the $\mathbb{R}^3$ symmetry present in $\Lambda$CDM \cite{Cea:2014gga,Cea:2019gnu}, which is exactly realised in Bianchi VII$_0$.

Let us begin this section by briefly discussing the standard analysis of the CMB temperature anisotropies~\cite{Campanelli:2007qn, Cea:2014gga}. The temperature anisotropy is given below:
\begin{equation}\label{eq:teman1}
\frac{\Delta T}{T}(\theta, \phi)=\frac{T(\theta, \phi)-\langle T\rangle}{\langle T\rangle}
\end{equation}
where $\theta$ and $\phi$ are angular coordinates on the celestial sphere (analogous to latitude and longitude on the surface of the Earth). The temperature anisotropy \eqref{eq:teman1} can be expanded in spherical harmonics:

\begin{equation}\label{eq:Ylmsum}
\frac{\Delta T(\theta, \phi)}{\langle T\rangle}=\sum_{l=1}^{\infty} \sum_{m=-l}^l a_{l m} Y_{l m}(\theta, \phi) .
\end{equation}
where $Y_{\ell}^m(\theta, \phi)$ are the usual spherical harmonic functions
\footnote{\begin{equation}
\begin{aligned}
Y_0^0(\theta, \phi) & =\frac{1}{2} \sqrt{\frac{1}{\pi}}, \, Y_1^{-1}(\theta, \phi) =\frac{1}{2} \sqrt{\frac{3}{2 \pi}} \sin \theta e^{-i \phi}, \, Y_1^0(\theta, \phi) =\frac{1}{2} \sqrt{\frac{3}{\pi}} \cos \theta \\
Y_1^1(\theta, \phi) & =-\frac{1}{2} \sqrt{\frac{3}{2 \pi}} \sin \theta e^{i \phi}
\end{aligned}
\end{equation}
and so on.}.
A statistical measure of the temperature fluctuations is the correlation function, $C(\theta)$. Consider two points on the surface of last scattering, in directions represented by the vectors $\mathbf{r}$ and $\mathbf{r}^{\prime}$, separated by the angle $\theta$ such that $\cos \theta=\mathbf{r} \cdot \mathbf{r}^{\prime}$. The correlation function $C(\theta)$ is found by multiplying together the values of $\Delta T / T$ at the two points, and averaging the product over all pairs of points separated by the angle $\theta$ :
\begin{equation}
C(\theta)=\left\langle\frac{\Delta T}{T}(\mathbf{r}) \frac{\Delta T}{T}\left(\mathbf{r}^{\prime}\right)\right\rangle_{\mathbf{r} \cdot \mathbf{r}^{\prime}=\cos \theta}
\end{equation}

\begin{equation}
C_{\ell}=\left\langle\left|a_{\ell}^m\right|^2\right\rangle=\frac{1}{2 \ell+1} \sum_m\left|a_{\ell}^m\right|^2 .
\end{equation}

In this way, the correlation function $C(\theta)$ can be broken down into its multipole components $C_{\ell}$.
The CMB temperature fluctuations are fully characterized by the power spectrum:
$$
\left(\frac{\Delta T_l}{\langle T\rangle}\right)^2 =\frac{\ell(\ell+1)}{2 \pi} C_{\ell}, \qquad C_{\ell}=\frac{1}{2 \ell+1} \sum_{m=-\ell}^{+\ell}\left|a_{\ell m}\right|^2 ;
$$
in particular, the quadrupole anisotropy refers to the multipole $\ell=2$.
Using the standard decomposition of the spherical harmonics in terms of the Legendre polynomial and using the orthonormality property one can rewrite the power spectrum:
\begin{equation}
\left(\frac{\Delta T_l}{\langle T\rangle}\right)^2=\frac{1}{2 \pi} \frac{l(l+1)}{2 l+1} \sum_m\left|a_{l m}\right|^2,
\end{equation}
that fully characterizes the properties of the CMB anisotropy. In particular, the quadrupole anisotropy refers to the multipole $\ell=2$ :
\begin{equation}
\mathcal{Q}^2 \equiv \left(\frac{\Delta T_2}{\langle T\rangle}\right)^2
\end{equation}
where $\langle T\rangle \simeq 2.7255 \mathrm{~K}$ is the actual (average) temperature of the CMB radiation. The Planck 2018 data \cite{Planck:2018vyg} determined that the observed quadrupole anisotropy is approximately $(\Delta T_2)^2\simeq 225.9~ \mu\text{K}^2$, whereas the best-fit values from the TT+TE+EE+low E+lensing under the assumption of the $\Lambda$CDM model gives $(\Delta T_2^I)^2 = 1017 \pm 643~ \mu\text{K}^2$, where the large errors are due to the effect of cosmic variance (and where we have added a superscript I for ``isotropic''). 

It has been proposed in \cite{Cea:2014gga,Cea:2019gnu,Cea:2022zep} (and others) that metric anisotropies may lower the quadrupole anisotropy to bring the theoretical best-fit more in line with the observed value, and we investigate here the consequences of our model on the CMB. If we consider that there is a small amount of anisotropy in the large scale spatial geometry of our Universe, then the observed CMB anisotropy map is a linear superposition of two contributions  :
\begin{equation}\label{eq:alms}
\Delta T=\Delta T_{\mathrm{A}}+\Delta T_{\mathrm{I}}
\end{equation}
where $\Delta T_{\mathrm{A}}$ is the contribution from the ansiotropic deviation of the geometry, while $\Delta T_{\mathrm{I}}$ is the standard isotropic contribution at the last scattering surface. The spherical harmonic coefficients $a_{l m}$  can be written as the summation of the contribution from both the isotropic and the anisotropic parts as below, 
\begin{equation}
a_{l m}=a_{l m}^{\mathrm{A}}+a_{l m}^{\mathrm{I}} .
\end{equation}
We are mostly interested in deriving the contribution to the deviation of the CMB radiation as a result of deviation of the geometry from the standard FLRW
 described by the metric \eqref{eq:metric}. We are working in the regime where the anisotropy is small. Considering the null geodesic equation we get that a photon emitted at the last scattering surface having energy $E_{\text {dec }}$ reaches the observer with an energy equal to $$E_0(\widehat{n})=\left\langle E_0\right\rangle\left(1-e_{\mathrm{dec}}^2 n_3^2 / 2\right),$$ where $\left\langle E_0\right\rangle \equiv E_{\mathrm{dec}} /\left(1+z_{\mathrm{dec}}\right)$, $e_{\rm dec}$ is the metric eccentricity (anisotropy) at the last scattering surface, and $\widehat{n}=\left(n_1, n_2, n_3\right)$ are the direction cosines of the null geodesic in the isotropic limit of the metric.
It is worthwhile to mention that the above result is derived for the case the case of the axis of symmetry directed along the $z$-axis. However this results can be easily generalised to the case  where the symmetry axis is directed along an arbitrary direction in a coordinate system $\left(x_{\mathrm{g}}, y_{\mathrm{g}}, z_{\mathrm{g}}\right)$ in which the $x_{\mathrm{g}} y_{\mathrm{g}}$-plane is the galactic plane. One can easily perform a rotation along the symmetry axis to derive a most generic result, where the axis are oriented along a general direction defined by the polar angles $(\vartheta, \varphi)$. Therefore, the temperature anisotropy in this new reference system is:
\begin{equation}
\begin{aligned}
\frac{\Delta T_{\mathrm{A}}}{\langle T\rangle} &\equiv \frac{E_0\left(n_{\mathrm{A}}\right)-\left\langle E_0\right\rangle}{\left\langle E_0\right\rangle}=-\frac{1}{2} e_{\mathrm{dec}}^2 n_{\mathrm{A}}^2 \\
n_{\mathrm{A}}(\theta, \phi)&=\cos \theta \cos \vartheta-\sin \theta \sin \vartheta \cos (\phi-\varphi).
\end{aligned}
\end{equation}
When the anisotropy is small, \eqref{eq:metric} may be written in a more standard form:
\begin{equation}
d s^2=-d t^2+a^2(t)\left(\delta_{i j}+h_{i j}\right) d x^i d x^j,
\end{equation}
where $h_{i j}$ is the metric perturbation which takes on the form:
\begin{equation}
h_{i j}=e(t)^2 \delta_{i 3} \delta_{j 3},
\end{equation}
and as a result, we can write the temperature anisotropies in a perturbed Friedmann-Lemaitre-Robertson-Walker through the null geodesic equation as (this is the integrated Sachs-Wolfe effect \cite{Nishizawa:2014vga}):
\begin{equation}
\frac{\Delta T}{\langle T\rangle}=\frac{1}{2} \int_{t_{\mathrm{dec}}}^{t_0} d t \frac{\partial h_{i j}}{\partial t} n^i n^j
\end{equation}
where $n^i$'s are the direction cosines.

In order to proceed, we need to determine the anisotropic spherical-harmonic expansion coefficients $a_{lm}^{\rm A}$ in Eq. \eqref{eq:alms}, which involves finding the temperature contrast function in terms of photon momentum through large-scale solutions of the Boltzmann equation. This has been worked out in great detail in \cite{Cea:2014gga}, and we present the main results here\footnote{for an exhaustive derivation of the temperature anisotropy we refer the readers to \cite{Cea:2014gga}.}.
The anisotropic contribution to the temperature contrast reads
\begin{equation}
    \frac{\Delta T_2^{\rm A}}{\langle T \rangle} = \frac{8\pi}{15}\theta_a\sum_{m=-2}^{+2}Y_{2m}(\theta,\phi)Y_{2m}^*(\theta_n,\phi_n),
\end{equation}
where $\theta_a$ is related to the degree of anisotropy at the surface of last scattering, and $\{\theta_n, \phi_n\}$ are the polar angles of the direction of $n^i$. We immediately find from Eq. \eqref{eq:Ylmsum} that
\begin{equation}\label{eq:almdeltae}
    a_{2m}^{\rm A} \simeq -\frac{4\pi}{15}\Delta e^2 Y^*_{lm}(\theta_n,\phi_n), \quad \Delta e^2 \equiv 0.944\left(e_{\rm dec}^2-e_0^2\right).
\end{equation}
This is where our analysis differ from that of \cite{Cea:2014gga}, where the authors assume that no anisotropy survives to the present day, and thus set $e_0=0$. In the equation above, we have reintroduced it, and the result is a shift in the spherical harmonic expansion coefficients which arises when solving for large-scale solutions of the Boltzmann equation. We find explicitly that
\begin{equation}\label{eq:almA}
    \begin{aligned}
        a_{20}^{\rm A} &\simeq \frac{\Delta e^2}{6}\sqrt{\frac{\pi}{5}}\left[1+3\cos^2{2\theta_n}\right] \\
        a_{21}^{\rm A} &= (a_{2,-1})^* \simeq i \sqrt{\frac{\pi}{30}}\Delta e^2e^{-i\phi_n}\sin{2\theta_n} \\
        a_{22}^{\rm A} &= (a_{2,-2})^* \simeq \sqrt{\frac{\pi}{30}}\Delta e^2 e^{-2\phi_n}\sin^2{\theta_n}. 
    \end{aligned}
\end{equation}
We can now define the anisotropic contribution to the total quadrupole anisotropy as
\begin{equation}
    \mathcal{Q}_{\rm A}^2 \equiv \left(\frac{\Delta T_2^A}{T_0}\right)^2 \quad \Longrightarrow \quad \mathcal{Q}_A \simeq \frac{2}{5\sqrt{3}}\Delta e^2,
\end{equation}
and we can find the explicit value for $\mathcal{Q}_A$ by plugging in our numerical solutions, and we find $\mathcal{Q}_A \sim 10^{-8}$.

We also need to determine the isotropic coefficients $a_{lm}^{\rm I}$, which necessarily need to respect $a_{l,-m} = (-1)^m(a_{l,m})^*$, since temperature anisotropies are real functions. This relation holds in the anisotropic case \eqref{eq:almA} and so must also hold for $a_{lm}^{\rm I}$. Furthermore, temperature fluctuations produced by standard inflation are statistically isotropic, so we take the same approach as in \cite{Cea:2014gga} and assume that the $a_{lm}^{\rm I}$'s are equal up to phase factors as
\begin{equation}
    \begin{aligned}
        a_{20}^I &\simeq \sqrt{\frac{\pi}{3}}\mathcal{Q}_I, \\
        a_{21}^{\rm I} &= -(a_{2,-1}^{\rm I}) \simeq i\sqrt{\frac{\pi}{3}}e^{i\phi_1}\mathcal{Q}_I \\
        a_{22}^{\rm I} &= (a_{2,-2}^{\rm I}) \simeq \sqrt{\frac{\pi}{3}}e^{i\phi_2}\mathcal{Q}_I,
    \end{aligned}
\end{equation}
where $\phi_1$ and $\phi_2$ are unknown phase factors, and the total coefficients are thus formed as $a_{lm} = a^I_{lm}+a^A_{lm}$. Finally, we find for the total quadrupole
\begin{equation}\label{eq:Qtot}
    \begin{aligned}
        \mathcal{Q}^2 &=\mathcal{Q}^2_{\rm I}+\mathcal{Q}^2_{\rm A} + 2f(\theta_n,\phi_n,\phi_1,\phi_2)\mathcal{Q}_{\rm I}\mathcal{Q}_{\rm A} \\
        f(\theta_n,\phi_n,\phi_1,\phi_2) &= \frac{1}{4\sqrt{5}}\left[1+3\cos{2\theta_n}\right]+\sqrt{\frac{3}{10}}\sin{2\theta_n}\cos{(\phi_2+\phi_n)} \\& +\sqrt{\frac{3}{10}}\sin^2{\theta_n}\cos{(\phi_2+2\phi_n)},
    \end{aligned}
\end{equation}
where the third term is a type of cross term. As such, it is possible that the total quadrupole anisotropy may become smaller than what is expected from standard $\Lambda$CDM. Due to the presence of the cross term, the phases present in the isotropic expansion coefficients acquire physical meaning. In the isotropic case, the unknown phases $\phi_1$ and $\phi_2$ are irrelevant, but this is no longer the case when anisotropy is non-zero. 

The Plank Collaborations are confirming the CMB anisotropies attributed to 
Lambda Cold Dark Matter model to the highest level of accuracy. However, at large angular scales there are still anomalous features in CMB anisotropies. One of the most evident discrepancy resides in the quadrupole $\langle TT\rangle$ correlation. The latest observed quadrupole $\langle TT\rangle$ correlation is:
$$
\left(\Delta T_2^{T T}\right)^2=225.90_{-132.37}^{+533.06} \mu K^2
$$
where the estimated errors take care of the cosmic variance. On the other hand, the 'TT,TE, EE + low E + lensing' best fit $\Lambda \mathrm{CDM}$ model to the Planck 2018 data gave:
$$
\left(\Delta T_2^{T T}\right)_{\Lambda C D M}^2=1016.73 \mu K^2
$$
that differs from the observed value by about two standard deviations.
Now if one assumes that the there is small amount of anisotropy in the geometry of our universe  then the quadrupole amplitude can be significantly reduced without affecting higher multipoles of the angular power spectrum of the temperature anisotropies \cite{Campanelli:2007qn,Campanelli:2006vb}.

\begin{figure}[t]
    \centering
    \caption{The off-diagonal angular function $f(\theta_n,\phi_n,\phi_1,\phi_2)$ (in orange), which needs to  lie below $-\mathcal{Q}_A/(2\mathcal{Q_I})$ (in blue). Shown here with the symmetry axis as found in \cite{Cea:2014gga,Cea:2019gnu}.}
    \includegraphics[scale=1]{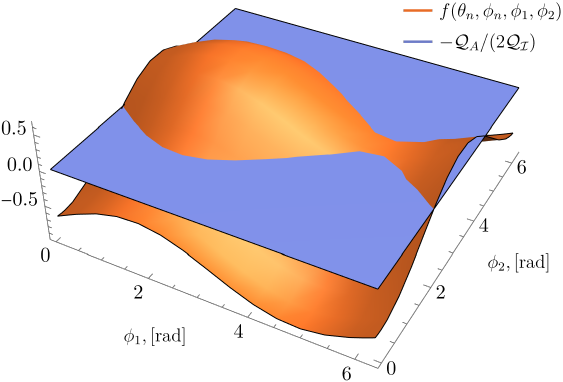}
    \label{fig:fDiscr}
\end{figure}

In order to solve or improve the quadrupole anomaly, the following relation must hold
\begin{equation}\label{eq:Qrel}
    \frac{\mathcal{Q}_A}{\mathcal{Q_I}}+2f(\theta_n,\phi_n,\phi_1,\phi_2) < 0,
\end{equation}
which can be read from Eq.~\eqref{eq:Qtot}. In order for this relation to hold, the function $f(\theta_n,\phi_n,\phi_1,\phi_2)$ must be negative and  $f(\theta_n,\phi_n,\phi_1,\phi_2)<-\mathcal{Q}_A/(2\mathcal{Q_I})$. We can now manipulate the symmetry axis $\{\theta_n,\phi_n\}$ and the phases $\{\phi_1,\phi_2\}$ in order to satisfy this relation, and through numeric manipulation we find that for every choice of symmetry axis, one can tune the phases such that Eq.~\eqref{eq:Qrel} holds, i.e. the quadrupole anisotropy is reduced. We pick a specific symmetry axis which coincides with that found in \cite{Cea:2014gga,Cea:2019gnu}: $\{\theta_n=73^\circ,\phi_n=264^\circ\}$, and we plot the function $f(\theta_n,\phi_n,\phi_1,\phi_2)$ along with $-\mathcal{Q}_A/(2\mathcal{Q_I})$ in Figure~\ref{fig:fDiscr}.
We see from this figure that for this choice of symmetry axis, the majority of the $\{\phi_1,\phi_2\}$ parameter space improves the quadrupole anomaly. We can now investigate how much it can be improved for different values of the phase angles. Keeping to the symmetry axis we chose above we evaluate Eq.~\eqref{eq:Qtot}, and we find that $\mathcal{Q}^2\sim 1.3764\cdot 10^{-10}$ for $\Theta=0$ and $\Theta=5\cdot10^8$, the difference being on the order of $10^{-8}$. Given this, we are able to reduce the quadrupole anomaly on the order of $0.39\%$. Another option is to {\it not} plug in a symmetry axis a priori, and instead minimize $\mathcal{Q}^2$ in Eq. \eqref{eq:Qtot} directly\footnote{For which we employ the function \texttt{NMinimize} in Wolfram Mathematica.}, which yields the optimal symmetry axis for our model as $\{\theta_n=291^\circ, \phi_n=47^\circ\}$ along with a small increase in the reduction of the quadrupole anomaly; we manage to reduce it by $\sim 0.4\%$, and thus our model is not successful in reducing the anomaly. The reason for this can be read off from Eq. \eqref{eq:almdeltae}, where the remaining anisotropy at the present time reduces the value of $\Delta e^2$, and thus that of $\mathcal{Q}_A$; the difference between $\Theta=0$ and $\Theta=5\cdot 10^8$ here is on the order of $10^{-8}$ (and therefore negligible).

\subsection{Dark Energy EOS}
We can analyse the anisotropic contribution to the dark sector by attributing it to dynamical dark energy. For that purpose, we can write the anisotropic stress-energy tensor \eqref{anstr} in the standard form as 
\begin{equation}
    T_{\mu\nu}^{\rm AN} = 
\begin{pmatrix}
    \rho^{\rm AN} & 0 & 0 & 0 \\
    0  & & & \\
    0 &  & g_{ij} p^{\rm AN}_i &  \\
    0 & & &
\end{pmatrix}.
\end{equation}
In the isotropic and homogeneous cosmological models we 
 can assume an equation of state of the form
\begin{align}
P=\omega \rho,
\end{align}
but in the presence of anisotropic matter sources and anisotropy induced in the geometry, the total pressure and the total energy density can similarly be split into isotropic and anisotropic parts as
\begin{equation}
\begin{aligned}\label{eq:rhopAN}
    \rho_t &= \left(\rho^{\rm PF}+\rho_0^{\rm AN}\right)+\epsilon \rho_1^{\rm AN},\\
    P_t &= \left( P^{\rm PF}+P_i^{\rm AN(0)}\right)+\epsilon P_i^{\rm AN(1)}.
\end{aligned}
\end{equation}
We can determine the effective equation of state parameter $w_t$ for the cosmic fluid, as was also noted in \cite{Koivisto:2005mm,Koivisto:2008ig,Appleby:2012as,Appleby:2009za,Guarnizo:2020pkj}. We have explicitly shown in \cite{Lee:2022rtz} that the perfect-fluid part also receives corrections at order $\epsilon$; these contributions are coupled to the anisotropic degrees of freedom, and we count them as part of $\rho_1^{\rm AN}$ and $P_i^{\rm AN}$.

\begin{figure}
\captionsetup[subfigure]{position=b}
     \centering
        \caption{The total equation of state parameter $w_{\rm tot}$.}
         \subcaptionbox{The total equation of state parameter $w_{\rm tot}$ for three different values of the SU(2) coupling constant $\Theta$\label{fig:wtot}}{\includegraphics[width=0.48\linewidth]{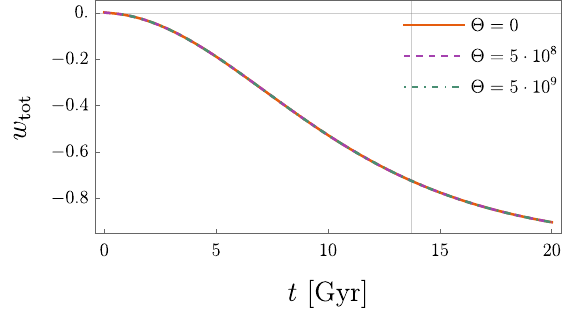}}
         \hfill
         \subcaptionbox{The total equation of state parameter $w_{\rm tot}$ at early time for three different values of the SU(2) coupling constant $\Theta$\label{fig:wtotearly}}{\includegraphics[width=0.48\linewidth]{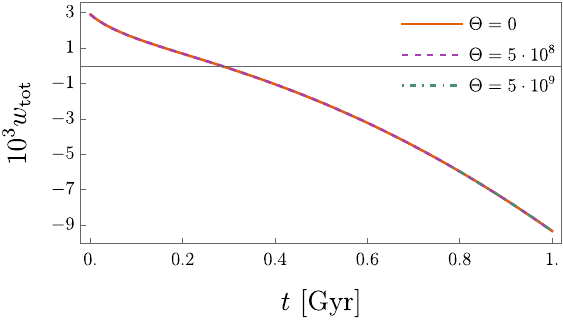}}
        \label{fig:wtotfull}
\end{figure}

\begin{figure}[h]
    \centering
    \caption{Ratios of the equation of state parameter $w_{\rm tot}$ for three different values of the SU(2) coupling constant $\Theta$}
    \includegraphics[scale=0.8]{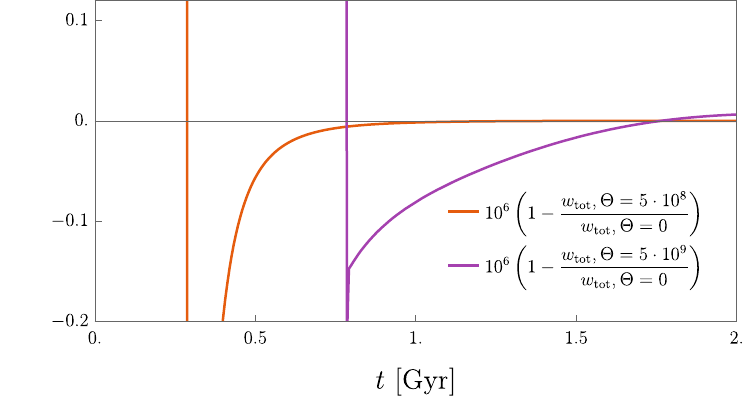}
    \label{fig:wtotratio}
\end{figure}

We identify the total energy density $\rho_{\rm tot}$ and pressure $p_{\rm tot}$ from the first and second Friedmann equation ((00) and (ii) components of the Einstein equations) and form the equation of state as $w_{\rm tot} = p_{\rm tot}/\rho_{\rm tot}$, and we plot this quantity for different values of $\Theta$ in Figures~\ref{fig:wtot}-~\ref{fig:wtotearly}, as well as the ratio between different values of $\Theta$ in Figure~\ref{fig:wtotratio}.
In Figure~\ref{fig:wtot} we note that our model evolves smoothly from $w_{\rm tot}\approx 0$ at early times and approaches $w_{\rm tot} \to -1$ at late times, i.e. the Universe is matter dominated at early times, and evolves smoothly to a $\Lambda$-dominated state.
It would seem that a radiation era ($w_t=1/3$) is missing, but Figure~\ref{fig:wtotearly} reveals that the equation of state crosses zero at $t\approx 0.28$ Gyr, approaching $w_{\rm ot}\approx 1/3$; this behaviour persists for all values of $\Theta$.
As for the other quantities studied in this section, we inspect the ratio of the equation of state for different values of $\Theta$, which can be seen in Figure~\ref{fig:wtotratio}. Here we see that there are significant differences at early times\footnote{Although the divergences simply indicates that the $\Theta=0$ solution crossed from negative to positive values and thus has no physical meaning}, before decaying to very similar values after $t=2$ Gyr. We note that in contrast to the other quantities, the most interesting features of the total equation of state seem to appear for $t<1$ Gyr.

\section{Discussion and Conclusions}\label{sec:concl}
In this paper we introduced a non-abelian version of the model presented in \cite{Lee:2022rtz} which induces consmological anisotropies in the geometry as a consequence of the backreaction of the matter sector. We choose the components of the gauge field $A_\mu$ to be aligned with the Killing vectors of the Bianchi VII$_0$ metric, and it was explicitly shown in \cite{Lee:2022rtz} the gauge field satisfies the same isometries as Bianchi VII$_0$.
We use a similar methodology as outlined in \cite{Lee:2022rtz} to solve the coupled set of differential equations using a perturbative scheme. The resulting system of equations are solved numerically and we recover the canonical 
$\Lambda$CDM solutions at zeroth order, with anisotropic contributions appearing at first order. Owing to the non-trivial parametrization of the gauge field, we obtain solutions to the anisotropic scale factors $\beta^{(1)}_i$ which are driven by the evolution of the gauge field $A_\mu$, and from the explicit solutions of the average Hubble parameter $\bar{H}$, we see that the deviation from $\Lambda$CDM is largest in the early universe, before settling down to the asymptotic (attractor) FLRW fixed point.

One of the most interesting feature of the non-abelian case as compared to its abelian cousin is the possibility to shed some light on the cosmological constant problem, which states that the evolution of the Universe should have lower value of the cosmological constant (ideally zero) as compared to the $\Lambda$CDM model \cite{Adler:1995vd,Bengochea:2019daa}. One could qualitatively represent the potential in \ref{action1} as some form of a field-dependent cosmological constant, $\Lambda(\phi)$, where the lower the value of such a potential would have implications for the cosmological constant problem. 

As has been noted in \cite{Lee:2022rtz}, the  magnitude of $\bar{H}$ is always smaller than $H_{\Lambda \rm CDM}$, and a negative slope at all times, which may have implications for the Hubble tension.  It is worthwhile to note here that the isotropic scale factor exhibits approximately standard $\Lambda$CDM evolution throughout the history of the Universe, although the amplitude is consistently higher; this is an artefact of our choice to impose initial conditions at $t=20$ Gyr. Our solutions for the anisotropic scale factors exp$(\beta^{(1)}_1)$ and exp$(\beta^{(1)}_2)$ are very similar in amplitude, but not identical; this is a desirable feature, since cosmological anisotropies are expected to be small, and by evaluating exp$(\beta^{(1)}_1)$ and exp$(\beta^{(1)}_2)$ at the present time ($t_0=1/H_0$), we find that the anisotropic expansion is on the order of $10^{-7} - 10^{-8}$; by examining $\bar{H}$ in Figure~\ref{fig:Haverage}, we see that a large part of the anisotropies have decayed away at $t=5$ Gyr. The scalar field $\phi$ exhibits steep falloff in the early Universe and settles down to a small constant at late times, and we find similar behaviour in $\psi$ and $\sigma$, which parameterize the gauge field. A related model was studied in \cite{Watanabe:2009ct} and similar results were found, but as discussed in the introduction, this is gauge-inequivalent to our model. 
We have also compared the average Hubble parameter for the $\Theta=0$ and $\Theta\neq0$ models in \ref{fig:hbarratio}. The average Hubble parameter of the $\Theta\neq0$ model is greater than the abelian cousin for different values of $\Theta$. This observations might be used as a differentiating diagnostic tool to analyze if the cosmological models beyond standard $\Lambda$CDM are closer to a abelian or non-abelian nature. One naive implication of the anisotropies induced by non-abelian gauge field would be the potential improvement of the Hubble tension, since the Hubble parameter is greater in the non-abelian case\footnote{This implies that the non-trivial interactions are important when proposing a model which would shed some light on the Hubble tension, but which we defer for future studies.}.
Taken together, these results indicate that most non-trivial effects will be contained in the early Universe. Whilst this does safeguard late-time evolution against large anisotropic effects, this is not necessarily desirable, since early-Universe processes (inflation, Big-Bang Nucleosynthesis (BBN), recombination etc) are very sensitive to the field content and initial conditions; in particular, early-Universe observables such as the sound horizon may be modified in the presence of anisotropies, in an analogous way to that of early dark energy \cite{Kamionkowski:2022pkx}.However, this lies beyond the scope of the present work. For studies regarding anisotropies in the inflationary era, see for example \cite{Watanabe:2009ct,Dulaney:2010sq,Gumrukcuoglu:2007bx,Gumrukcuoglu:2010yc,Pitrou:2008gk}.

In Appendix E of \cite{Lee:2022rtz} we have shown explicitly that the perfect-fluid part of the total stress-energy tensor receives anisotropic corrections perturbatively, both in the energy density and in the pressure. We also find off-diagonal components to the stress-energy tensor, which act as constraint equations, as was also studied in~\cite{Cho:2022xku}. The anisotropic part of the energy density has been studied as \emph{anisotropic dark energy}, for example in \cite{Koivisto:2005mm} and \cite{Koivisto:2008ig}, although at the background level. There are also interesting connections to the quadrupole anomaly in the CMB \cite{Rodrigues:2007ny}. 
The most important result of this work is the dynamical driving of cosmological anisotropies; we have shown that it is possible to find solutions which closely resemble those of $\Lambda$CDM at zeroth order, whilst containing a small degree of anisotropic correction at order $\epsilon$. An important note is that we are likely overestimating the magnitude of the dark-energy density $\Omega_\Lambda$: since the extra field content $\{\phi(t),\psi(t),\sigma(t),\beta_1(t),\beta_2(t)$ can be interpreted as dynamical dark energy, the total dark-energy density should read $\Omega_{\rm DE}=\Omega_{\Lambda}+\Omega_{\phi} + \hdots$, but because of the small scales of the anisotropies and the field $\phi(t)$, this would be a very small correction\footnote{For a discussion of the current observational status of dynamical dark energy, see \cite{SolaPeracaula:2018wwm}.}.

It has been advocated in several papers \cite{Cea:2014gga,Svrcek:2006yi,Cea:2022zep} that in order to reconcile the observed data on the quadruple correlation with the theoretically predicted values we need a small amount of anisotropy in the geometry. We have also shown that our model does not significantly alter the temperature quadrupole anisotropies preferred by the Planck data, which is a desirable result in anisotropic cosmology. By allowing a small deviation from FLRW geometry at the time of decoupling, a lower value of the temperature quadrupole can be generated, and can indeed be matched to the total quadrupole anisotropy by accounting for the unknown phases which are irrelevant in the isotropic limits, but which become physical as cross-terms in the presence of anisotropy. We note, however, that our model is not able to reconcile the Planck 2018 best fit to the observed value temperature quadrupole anisotropies. We leave for future work are more careful data analysis to compare this model with Planck data, as was done for the ellipsoidal Universe in \cite{Cea:2019gnu}.

As we see in the solutions, the anisotropic effects are larger in the early Universe before decaying and reaching the homogeneous and isotropic fixed point in the future, in keeping with the cosmic no-hair theorem; hence, any sizeable anisotropic expansion in the early Universe should affect the large-scale structure formation. We could perform a cosmological perturbation analysis of our model and thus get some hints about whether the anisotropic expansion is intertwined with the formation of large-scale structure. We leave such explorations for future studies.

On the observational side, the status of anisotropic cosmology is evolving, with tantalising results such as anisotropic acceleration (anomalous bulk flow) in the direction of the CMB dipole at $3.9\sigma$ significance~\cite{Colin:2019opb} and a $3\sigma$ hemispherical power asymmetry in the Hubble constant, also aligned with the CMB dipole\footnote{A possible solution to the hemispherical power asymmetry was recently proposed in \cite{Kumar:2022zff}.}~\cite{Luongo:2021nqh}. There are also hints of a preferred symmetry axis in the Pantheon+ sample of supernovae Type Ia \cite{McConville:2023xav}. Indications of fine structure-constant variation along with preferred directions in the CMB results in compelling evidence that the cosmological standard model is in need of revision, and in this paper we have provided a mechanism through which such preferred directions can arise from a well-motivated field theory. This is of course not the only model which can generate cosmological anisotropies; in particular, models exhibiting spacetime-symmetry breaking are known to contain preferred directions. For example, Ho\v{r}ava-Lifshitz gravity \cite{Horava:2009uw} Einstein-Aether theory \cite{Gasperini:1987nq}, and  bumblebee gravity~\cite{Maluf:2021lwh}, all of which have received attention in recent years, contain preferred frames. On the other hand, spacetime-symmetry breaking in gravity has been tightly constrained (see for example~\cite{Kostelecky:2008ts}). Our construction has the advantage of keeping these well-tested spacetime symmetries intact, and instead postulating the existence of new fields.

\section*{Acknowledgements}
     BHL thanks APCTP and KIAS for the hospitality during his visit, while part of this work has been done. BHL, WL and HL were supported by the Basic Science Research Program through the National Research Foundation of Korea (NRF) funded by the Ministry of Education, Science and Technology (BHL, HL: NRF-2020R1A6A1A03047877, BHL: NRF-2020R1F1A1075472, WL: NRF-2022R1I1A1A01067336, HL: NRF-2023R1A2C200536011). NAN was financed by CNES and IBS under the project code IBS-R018-D3, and acklowledges support from PSL/Observatoire de Paris. The work of ST was supported by Mid-career Researcher Program through the National Research Foundation of Korea grant No.\\ NRF-2021R1A2B5B02002603.

\vspace{1cm}
\appendix
\section{$SU(2)$ algebra  }\label{app:su2al}
In this appendix we give a brief outline of the notations and the basics of the  $SU(2)$ algebra that we used in the text.
Any non-abelian group has an $S U(2)$ subgroup. The gauge fields $A_i$ is in vector (triplet) representation of the rotation group $S O(3)_R$.
As far as our current discussion is concerned, without loss of generality, we can choose the gauge group $G$ to be $S U(2)$ or $S O(3)$ and choose the $T^A$'s to be $S U(2)$ generators in the triplet (adjoint) representation $\frac{1}{2} \sigma^a, a=1,2,3$ where $\sigma^a$ are Pauli matrices. Let us choose the background to be
\begin{equation}
    A_i^a = \psi_i(t) e^a_i,
\end{equation}
so that out of 12 components of $A^a{ }_\mu$, nine are physical and three are gauge freedom, which may be removed by a suitable choice of gauge parameter; we can safely use the temporal gauge. Out of the nine physical gauge fields we find that for each color indices $a=1,2,3$ the defining equations are same. Essentially we have three independent defining equations for the gauge fields. We work with the temporal gauge, $A_0^a=0$. When we are considering non-abelian gauge fields we have two symmetries associated with the one form gauge fields, namely the spacetime symmetry and the internal symmetry. As explained above, the SU(2) sector can be written as $A_{\mu}=\textrm{Tr}(A^a_{\mu}T^a)$, $F_{\mu\nu}=\textrm{Tr}(F^a_{\mu\nu}T^a)$. In the main text, we have expanded the gauge field in the basis of the Killing symmetry of the geometry, i.e in terms of the isometries of Bianchi VII$_0$, while the internal symmetry of the gauge fields commutes with the spacetime symmetry.

\section{First-order equations of motion}\label{app:pert}
The zeroth-order equations are listed in Section~\ref{sec:zeroth}, and here we list the more lengthy first-order corrections; therefore, the full order reads schematically
\begin{equation*}
    \text{Full order} = \text{zeroth order} + \epsilon\Big[\text{first order}\Big].
\end{equation*}
The gauge-field equations read
\begin{equation}
\begin{aligned}
        0 &= \ddot{\psi}^{(1)} - 2 \psi^{(0)} \ddot{\sigma}^{(1)} + \psi^{(0)} \ddot{\beta}_1^{(1)} +\Big( \beta_1^{(1)} - 2 \sigma^{(1)} \Big) \ddot{\psi}^{(0)}+ 3 \dot{\alpha}^{(0)} \dot{\psi}^{(1)} \\& + \Big[ \psi^{(1)} + \Big( \beta_1^{(1)} - 2 \sigma^{(1)} \Big) \psi^{(0)} \Big] \ddot{\alpha}^{(0)}
          + 3 \Big( \beta_1^{(1)}-2 \sigma^{(1)} \Big) \dot{\alpha}^{(0)} \dot{\psi}^{(0)} \\& + 2 \Big[ \psi^{(1)} + \Big( \beta_1^{(1)} - 2 \sigma^{(1)} \Big) \psi^{(0)} \Big] (\dot{\alpha}^{(0)})^2
        + \Big( \dot{\beta}_1^{(1)} + 2 \dot{\beta}_2^{(1)} \Big) \dot{\psi}^{(0)} \\& + 2 \Big( \dot{\beta}_1^{(1)} + \dot{\beta}_2^{(1)} - 3 \dot{\sigma}^{(1)} \Big) \psi^{(0)} \dot{\alpha}^{(0)} - g_A \Theta (\psi^{(0)})^2 \dot{\phi^{(1)}}
        \\&
        - g_A \Theta \Big[ \psi^{(1)} + \Big( \beta_1^{(1)} + 2 \sigma^{(1)} \Big) \psi^{(0)} \Big] \psi^{(0)} \dot{\phi}^{(0)}- 4 \dot{\psi}^{(0)} \dot{\sigma}^{(1)} \\& + 2 g_A^2 \Big( 3 \psi^{(1)} + \beta_1^{(1)} \psi^{(0)} \Big) (\psi^{(0)})^2,
       \quad  (\mu=a=1),
\end{aligned}
\end{equation}

\begin{equation}
\begin{aligned}
        0 &= \ddot{\psi}^{(1)} + \psi^{(0)} \ddot{\sigma}^{(1)} +\psi^{(0)} \ddot{\beta}_2^{(1)} + \Big( \beta_2^{(1)} + \sigma^{(1)} \Big) \ddot{\psi}^{(0)} - g_A \Theta (\psi^{(0)})^2 \dot{\phi^{(1)}}\\& + \Big[ \psi^{(1)} + \Big( \beta_2^{(1)} + \sigma^{(1)} \Big) \psi^{(0)} \Big] \ddot{\alpha}^{(0)}
        + 3 \dot{\alpha}^{(0)} \dot{\psi}^{(1)} + 2 \dot{\psi}^{(0)} \dot{\sigma}^{(1)} \\& +3 \Big( \beta_2^{(1)}+\sigma^{(1)} \Big) \dot{\alpha}^{(0)} \dot{\psi}^{(0)} + 2 \Big[ \psi^{(1)} + \Big( \beta_2^{(1)} + \sigma^{(1)} \Big) \psi^{(0)} \Big] (\dot{\alpha}^{(0)})^2
        \\&
        + \Big( \dot{\beta}_1^{(1)} + 2 \dot{\beta}_2^{(1)} \Big) \dot{\psi}^{(0)} + \Big( \dot{\beta}_1^{(1)} + 3 \dot{\beta}_2^{(1)} + 3 \dot{\sigma}^{(1)} \Big) \psi^{(0)} \dot{\alpha}^{(0)}
        \\&
        - g_A \Theta \Big[ 2 \psi^{(1)} + \Big( \beta_2^{(1)} - \sigma^{(1)} \Big) \psi^{(0)} \Big] \psi^{(0)} \dot{\phi}^{(0)} \\& + 2 g_A^2 \Big( 3 \psi^{(1)} + \beta_2^{(1)} \psi^{(0)} \Big) (\psi^{(0)})^2,
       \quad (\mu=a=2, 3).
\end{aligned}
\end{equation}

The lengthy components of the decomposed Einstein equations read
\begin{equation}
    \begin{aligned}
        0 &= \ddot{\beta}_1^{(1)} + \ddot{\beta}_2^{(1)} + 2 \left( 2 \ddot{\alpha}^{(0)} + 3 (\dot{\alpha}^{(0)})^2 \right) \beta_2^{(1)} + 3 \dot{\alpha}^{(0)} \left( \dot{\beta}_1^{(1)} + \dot{\beta}_2^{(1)} \right)
        \\&
        + \kappa \Big[ \beta_2^{(1)} (\psi^{(0)})^4 g_A^2 + 2 \sigma^{(1)} (\psi^{(0)})^4 g_A^2 + 2 (\psi^{(0)})^3 \psi^{(1)} g_A^2 + \\& \dot{\alpha}^{(0)} \dot{\beta}_1^{(1)} (\psi^{(0)})^2
        + (\dot{\alpha}^{(0)})^2 \beta_2^{(1)} (\psi^{(0)})^2 + 2 \dot{\alpha}^{(0)} \beta_2^{(1)} \dot{\psi}^{(0)} \psi^{(0)} \\& - 2 (\dot{\alpha}^{(0)})^2 \sigma^{(1)} (\psi^{(0)})^2
        - 2 \dot{\alpha}^{(0)} \dot\sigma^{(1)} (\psi^{(0)})^2 - 4 \dot{\alpha}^{(0)} \sigma^{(1)} \dot{\psi}^{(0)} \psi^{(0)} \\& + (\dot{\alpha}^{(0)})^2 \psi^{(0)} \psi^{(1)} + \dot{\alpha}^{(0)} \psi^{(0)} \dot{\psi}^{(1)}
        + \dot{\alpha}^{(0)} \dot{\psi}^{(0)} \psi^{(1)} + \dot{\beta}_1^{(1)} \dot{\psi}^{(0)} \psi^{(0)} \\& + \beta_2^{(1)} (\dot{\psi}^{(0)})^2 + \beta_2^{(1)} (\dot{\phi}^{(0)})^2 + 2 \beta_2^{(1)} p
         - 2 \dot\sigma^{(1)} \dot{\psi}^{(0)} \psi^{(0)} - 2 \sigma^{(1)} (\dot{\psi}^{(0)})^2 \\& + \dot{\psi}^{(0)} \dot{\psi}^{(1)} + \dot{\phi}^{(0)} \dot{\phi}^{(1)} - 2 \beta_2^{(1)} V(\phi^{(0)}) \\& - \phi^{(1)} V'(\phi^{(0)}) \Big],
       \quad  (\mu=\nu=1),
    \end{aligned}
\end{equation}

\begin{equation}
    \begin{aligned}
        0 &= \ddot{\beta}_1^{(1)} + \ddot{\beta}_2^{(1)} + 2 \left( 2 \ddot{\alpha}^{(0)} + 3 (\dot{\alpha}^{(0)})^2 \right) \beta_2^{(1)} + 3 \dot{\alpha}^{(0)} \left( \dot{\beta}_1^{(1)} + \dot{\beta}_2^{(1)} \right)
        \\&
        + \kappa \Big[ \dot{\alpha^{(0)}} \dot{\beta}_1^{(1)} (\psi^{(0)})^2 + (\dot{\alpha^{(0)}})^2 \beta_2 (\psi^{(0)})^2 + 2 \dot{\alpha^{(0)}} \beta_2 \dot{\psi}^{(0)} \psi^{(0)} \\& - 2 (\dot{\alpha}^{(0)})^2 \sigma^{(1)} (\psi^{(0)})^2
        - 2 \dot{\alpha}^{(0)} \dot{\sigma}^{(1)} (\psi^{(0)})^2 - 4 \dot{\alpha}^{(0)} \sigma^{(1)} \dot{\psi}^{(0)} \psi^{(0)} \\& + (\dot{\alpha}^{(0)})^2 \psi^{(0)} \psi^{(1)} + \dot{\alpha}^{(0)} \psi^{(0)} \dot{\psi}^{(1)}
        + \dot{\alpha}^{(0)} \dot{\psi}^{(0)} \psi^{(1)} + \dot{\beta_1}^{(1)} \dot{\psi}^{(0)} \psi^{(0)} \\& + \beta_2^{(1)} (\dot{\psi}^{(0)})^2 + \beta_2^{(1)} (\dot{\phi}^{(0)})^2 + 2 \beta_2^{(1)} p 
        - 2 \dot{\sigma}^{(1)} \dot{\psi}^{(0)} \psi^{(0)} \\&- 2 \sigma^{(1)} (\dot{\psi}^{(0)})^2 + \dot{\psi}^{(0)} \dot{\psi}^{(1)} + \dot{\phi}^{(0)} \dot{\phi}^{(1)} - \phi^{(1)} V'(\phi^{(0)}) - 2 \beta_2^{(1)} V
        \\&
          + g_A^2 \left( (\psi^{(0)})^4 \beta_2^{(1)} + 2 (\psi^{(0)})^4 \sigma^{(1)} + 2 (\psi^{(0)})^3 \psi^{(1)} \right) \Big],
        \quad(\mu=\nu=2, 3).
    \end{aligned}
\end{equation}
In the  non-abelian model we notice that the off-diagonal elements of the Einstein equations vanish, which can be simply understood by the non-mixing of the color indices of the non-abelian gauge fields.

\newpage
\bibliographystyle{elsarticle-num}
\bibliography{Refs}

\end{document}